# A Deep Learning Approach for Modeling and Hindcasting Lake Michigan Ice Cover

**Hazem U. Abdelhady, Cary D. Troy***

Purdue University, Lyles School of Civil Engineering, 550 Stadium Mall Drive, West Lafayette, IN, 47907-2051, USA

* Corresponding author: troy@purdue.edu

## Abstract

In large lakes, ice cover plays an important role in shipping and navigation, coastal erosion, regional weather and climate, and aquatic ecosystem function. In this study, a novel deep learning model for ice cover concentration prediction in Lake Michigan is introduced. The model uses hindcasted meteorological variables, water depth, and shoreline proximity as inputs, and NOAA ice charts for training, validation, and testing. The proposed framework leverages Convolution Long Short-Term Memory (ConvLSTM) and Convolution Neural Network (CNN) to capture both spatial and temporal dependencies between model input and output to simulate daily ice cover at 0.1° resolution. The model performance was assessed through lake-wide average metrics and local error metrics, with detailed evaluations conducted at six distinct locations in Lake Michigan. The results demonstrated a high degree of agreement between the model's predictions and ice charts, with an average RMSE of 0.029 for the daily lake-wide average ice concentration. Local daily prediction errors were greater, with an average RMSE of 0.102. Lake-wide and local errors for weekly and monthly averaged ice concentrations were reduced by almost 50% from daily values. The accuracy of the proposed model surpasses currently available physics-based models in the lake-wide ice concentration prediction, offering a promising avenue for enhancing ice prediction and hindcasting in large lakes.

## Introduction

Ice cover in the Great Lakes plays an important role in the region's environment and economy. During the winter months, ice cover dynamics in the Great Lakes affect shipping (Millerd, 2011, 2005), weather (Benson et al., 2012; Deser et al., 2000), wave generation and coastal erosion (Anderson et al., 2015; BaMasoud and Byrne, 2012; Dodge et al., 2022), and the aquatic ecosystem (e.g. fish recruitment (Lynch et al., 2010; Magnuson et al., 1997). Ice cover is also directly linked to interannual and long-term variability of Great Lakes water levels, by regulating over-winter evaporation (Woolway et al., 2020). As climate change steadily reduces ice cover in the Great Lakes (Cannon et al., 2024; Jensen et al., 2007; Woolway et al., 2020), all of these processes will be affected, and our ability to understand and manage these changes is predicated in part on our abilities to model the dynamics of ice cover in the Great Lakes.

Physics-based numerical models have been used to simulate ice cover in the Great Lakes (Anderson et al., 2018; Wang et al., 2010). These models are based on coupling an ice model (e.g., CICE (Hunke et al., 2017); CIOM (Wang et al., 2002)) with a coastal ocean model (e.g., POM (Blumberg and Mellor, 2012); FVCOM (Chen et al., 2006; Wang et al., 2010)). Ice models usually include several interactive components such as thermodynamics, ice dynamics, ice transport, and ice deformation parameterization functions (Hunke et al., 2017), while coastal ocean models are used to provide the ice models with the heat and momentum fluxes, calculated using a 3D hydrodynamic model that is forced by atmospheric conditions like wind speed, air temperature, dew point, and cloud cover (Wang et al., 2010).

Despite their utility, ice models exhibit notable limitations and reported errors can be significant (Johnson et al., 2019; Orsolini et al., 2019). They often demand substantial computational resources (Craig et al., 2015), may suffer from numerical instability (Kiss et al., 2020; Lipscomb et al., 2007), and may not account for all the intricacies of ice formation in the Great Lakes (Anderson et al., 2018). Furthermore, the coupling between ice and coastal ocean models amplifies computational demands and instability risks.

More specifically for the Great Lakes, these models have more errors during the freezing and melting periods (Anderson et al., 2018; Wang et al., 2010). They usually have more errors in capturing the spatial distribution of the ice cover than the temporal changes of the lakewide averaged concentration with reported errors reaching up to 90% (Anderson et al., 2018; Wang et al., 2010). These discrepancies are attributed to the limitation of the current ice models in capturing some ice dynamics processes (e.g., ice-wave interaction and land-fast ice) and uncertainty in the meteorological forcing (Anderson et al., 2018).

Deep learning (DL) has emerged as a powerful tool in geophysics and climate research, revolutionizing our capacity to analyze vast and intricate datasets and unveil concealed patterns within Earth's systems (Yu and Ma, 2021). Deep learning models can also capture the nonlinear relation between the input and the output data without being limited to our understanding of the physics (Baldi et al., 2014). In the context of the Great Lakes region, DL applications have made notable strides, aiding in the prediction of wave heights in Lakes Michigan and Lake Erie (Feng et al., 2020; Hu et al., 2021) and surface water temperature assessments (Xue et al., 2022).

To date, the application of DL methods for ice prediction in the Great Lakes region has been limited to either water channels connecting the lakes (e.g. St. Mary's River, (Liu et al., 2022)) or smaller lakes (e.g., multiple lakes in Canada (Zaier et al., 2010)). However, DL methods have been applied with success to ice prediction in the oceans. Recently, Andersson et al., (2021) developed a deep learning model for Arctic ice forecasting that outperforms simple statistical benchmarks and physics-based models. However, there are differences between Arctic ice dynamics and the Laurentian Great Lakes. The alternation between ice-fast and ice-free seasons and the significant change in the water depth between the nearshore area and the middle of the lake may make ice dynamics in the Great Lakes more difficult to predict.

In this paper, a hybrid deep learning model, based on Convolution Long Short-Term Memory (ConvLSTM) and Convolution Neural Network (CNN) approaches, is introduced for ice cover concentration modeling and hindcasting in the Great Lakes. The model is developed for Lake Michigan, utilizing inputs derived from the ERA5-Land weather dataset (Muñoz Sabater, 2019), water levels (NOAA, 2021), and lake bathymetry data. Training, validation, and testing of the model were conducted using NOAA ice chart datasets spanning from 1973 to 2022. The model's performance was assessed using both lake-wide and local simulation metrics, examined at six different locations in Lake Michigan (Figure 1).

## Methods

### Model input (features)

In this study, the deep learning model utilized various datasets as input features, including meteorological data, water depth, water levels, and land proximity. The input data for the model was selected from a region spanning Lake Michigan (-88.6° to -84.8° longitudes and 41.0° to 46.4° latitudes). The initial choice of the input variables is based on the connection between these variables and the formation of ice in the Great Lakes. The meteorological data were drawn from ECMWF ERA5-Land (Muñoz Sabater, 2019), which operates at a spatial resolution of 0.1°. ERA5-Land's reanalysis variables are the result of data assimilation, a process that combines observational data from weather stations around the world with a physics-based atmospheric model to create a consistent gridded dataset adhering to the laws of physics and consistent with the observations (Muñoz Sabater, 2019).

The variables extracted from the ERA5-Land hourly surface dataset are the following: air temperature at 2 m elevation; net shortwave radiation; and wind speed at 10 m elevation in both the east and north directions, which was converted to wind speed and direction. These variables were averaged to daily averages for compatibility with the ice atlas used for model training and validation. Air temperature has been found to explain 70% of the variability in the ice cover in the Great Lakes (Titze, 2016), while net short-wave radiation and wind speed and direction also have been shown to have a large influence on ice cover (Austin and Colman, 2007; Wang et al., 2012, 2005). Variables that were originally included from ERA5 in the model as input features but later discarded due to their lack of effects on model results included dewpoint temperature, surface pressure, relative humidity, and longwave radiation. Also, other variations of the wind speed and direction, like wind speed in the x and y direction and wind speed with sine and cosine of wind direction, were tested and found to get worse results. Finally, some climate indices (e.g., North Atlantic Oscillation (NAO)), which were extracted from other climate datasets (e.g., NOAA Oceanic & Atmospheric Climate Data), were also tested and discarded due to their lack of effect on the model results.

Additionally, two more model input time series were created from 2 m air temperature, the trailing moving average temperature with a window of 30 and 150 days. In essence, these slowly-varying air temperature time series serve as surrogates for the lake water temperature, which is not simulated by the present model but is perhaps the most important variable influencing ice formation, especially for freshwater (Hunke et al., 2017). Previous research by Assel, (1990), (1976) and Wang et al., (2012) showed that ice cover has a strong relationship to air temperatures on monthly and seasonal timescales. This is mainly due to the large heat capacity of the Great Lakes especially due to the large water depth and the lake-wide circulation (Fink et al., 2014; Wang et al., 2010). Thus, these two air temperature features were added, in addition to the original 2m air temperature, and the two moving average periods were considered hyperparameters.

Lake water depth and distance from shore were also used as input features. Previous research has shown that different parts of the Great Lakes experience different ice cover based on water depth (Wang et al., 2012), and ice in the nearshore area are behaving differently due to the land-fast ice phenomenon (Anderson et al., 2018; Lin et al., 2022). Spatially- and temporally-variable water depths were generated by combining published bathymetric data for Lake Michigan (https://www.ncei.noaa.gov/products/great-lakes-bathymetry) with the daily water levels from the Calumet Harbor, IL water level gage (https://tidesandcurrents.noaa.gov/). The land proximity information was also included for each grid cell by calculating the shortest distance between each grid cell and the shoreline which was extracted from USGS Great Lakes and Watershed Shapefiles (USGS, 2010). All the cells onshore were given a zero value for the water depth and the land proximity. The lake water depth and distance from shore were resampled to match the cell grid used for the ERA5-Land variables. All input features used in the model are shown in Table 1.

### The ice cover charts dataset

Lake Michigan ice data for this study were obtained from the NOAA CoastWatch Great Lakes Node (NOAA, 2022). This dataset provides ice concentration information for the Great Lakes region, sourced from the U.S. National Ice Center (NIC). The ice analysis products in this dataset are derived from various satellite sources, including Radarsat-2, Envisat, AVHRR, Geostationary Operational and Environmental Satellites (GOES), and the Moderate Resolution Imaging Spectroradiometer (MODIS). The spatial resolution of the ice concentration data is 1.8 kilometers, as reported by (Yang et al., 2020). The NIC dataset represents ice concentration values ranging from 0 to 100% in 5% increments before 1983, and in 10% increments after 1983 (Yang et al., 2020), which represents the accuracy of the dataset. The lake-wide average ice chart dataset is shown in Figure 2.

To ensure compatibility with the spatial resolution of the input features, the ice chart dataset was coarsened to match ERA5-Land's spatial resolution by performing spatial averaging. The process of spatially averaging resulted in ice concentrations that were no longer restricted to the 5% and 10% intervals mentioned earlier.

### Hybrid ConvLSTM-CNN model

The ConvLSTM model was adapted to simulate ice cover concentrations in Lake Michigan. ConvLSTM combines the strengths of Convolutional Neural Networks (CNN) for capturing spatial dependencies and Long Short-Term Memory (LSTM) networks for modeling temporal dependencies (Shi et al., 2015). Given that ice cover concentration on a specific day is influenced not only by the meteorological conditions of that day but also by those of previous days and months, it was imperative to select a deep learning (DL) model capable of accounting for the temporal dependencies between input features and ice cover concentration. Additionally, as the ice cover in a particular grid cell is influenced by its surrounding conditions, such as proximity to land and heat flux from neighboring cells, it is desirable to leverage a DL model that can account for spatial variability and dependencies.

The ConvLSTM model, originally introduced by (Shi et al., 2015), emerged initially in the context of precipitation forecasting. It builds upon the concept of stacked LSTM models but introduces convolution operators within the state-to-state and input-to-state transitions. This innovation diverges from the standard Hadamard product typically used in stacked LSTMs, enabling the ConvLSTM to effectively address spatial correlations inherent in the data as well as the temporal correlations. For a comprehensive understanding of the ConvLSTM's key equations, please refer to the original paper (Shi et al., 2015).

For the present application to ice cover simulation, the ConvLSTM takes a 4D input (latitude x longitude x lookback time x features) and converts it to a 3D output (latitude x longitude x features) which is subsequently fed into multiple CNN layers, each with a varying number of features, to extract spatial features and map the input features to the ice cover concentration. The ultimate output, which represents the predicted ice cover concentration, is generated by a final CNN layer with one feature. The final model architecture is discussed later in the methods section. The model was developed and implemented in Python 3.10 using the TensorFlow deep learning library. All training procedures were conducted using an AMD EPYC 7702P 64-Core Processor, with each model training session taking approximately 10 hours.

### Model training and hyperparameter optimization

Both input features and output labels were normalized to improve the speed and stability of the model training. In DL, feature normalization is a standard procedure that ensures that the input features and output labels have comparable magnitudes, aiding in the optimization of the model parameters (Goodfellow et al., 2016). For the input features, the normalization process was performed by subtracting the mean value and dividing it by the standard deviation. For the output labels (ice cover concentration), which range from 0 to 100, the values were divided by 100 which resulted in output labels ranging between 0 and 1.

The input and output datasets were divided into 3 distinct datasets for training, validation, and testing (Figure 2). The training dataset spanned from 1973 to spring 2010 and served as the foundation for training the ConvLSTM models developed in this study. The validation dataset covered the period from the fall of 2010 to the spring of 2018 and was used in the early stopping criterion to avoid overfitting as explained later. The validation dataset was also used to evaluate the model performance after training during the search for the model's hyperparameters as explained later. Finally, the testing dataset from fall 2018 to spring 2022 was used as a blind testing phase to evaluate the model's performance and provide an unbiased assessment of the model's ability to predict ice cover concentration on unseen data.

Model training was carried out using complete ice seasons, spanning from October to the end of April to avoid the data imbalance concerns, as it excluded months with mostly zero ice cover. The models' weights were trained using backpropagation of a customized loss function based on the mean absolute error, explained later, with the Adam optimizer (Kingma and Ba, 2014). To prevent overfitting to the training dataset, model checkpointing and early stopping were implemented. In this method, the model performance on the validation dataset is monitored after each epoch, and the weights of the model with the best performance on the validation dataset are saved. The model training was stopped if the performance on the validation dataset did not improve, with patience of 10 epochs, and the weights of the best model were restored. This approach ensured that the network's weights, which yield the best generalization performance, were used for the final model, preventing performance degradation due to overfitting.

Numerous hyperparameters were optimized in the model, including the number of filters, number of hidden ConvLSTM layers, lookback period, number of CNN layers, kernel size, dropout ratio, activation function, the moving averaged periods for additional temperature features, and the loss function parameter α, explained later. To optimize these hyperparameters, we employed a dual-pronged approach, which combined grid search and manual adjustments. Manual refinements were made to the number of layers, lookback periods, and α, while systematic exploration, facilitated by for loops, was used to determine the optimal values for the remaining hyperparameters.

Ensemble learning and customized loss functions were employed to mitigate the bias in the model's predictions. The averaging ensemble learning approach was applied in this study, and it involved training 15 distinct models. Each model was initialized with random parameters and trained to predict daily ice cover based on the input features. Then, the loss on the validation dataset was calculated for all models, and the models were ordered based on their performance. Consequently, a series of ensemble models were created, ranging from 2 models to 15, by averaging the models' predictions. Finally, the best ensemble models on the validation dataset were used to get the final model prediction on the testing dataset.

A customized loss function that is based on the Mean Absolute Error (MAE) was used to better account for the imbalance in the training dataset. This loss function penalizes underestimation more than overestimation by introducing a tuning factor, denoted as α. α value ranges from 0.5 to 1 and was considered an additional hyperparameter that was tuned manually with the other hyperparameters. The loss function was defined as follows:

$$loss = \frac{1}{N_t N_x N_y} \sum_{t=1}^{N_t} \begin{cases} \sum_{i=1}^{N_x} \sum_{j=1}^{N_y} \alpha |y_{true(i,j)}^t - y_{pred(i,j)}^t| & if\ y_{true(i,j)}^t > y_{pred(i,j)}^t \\ \sum_{i=1}^{N_x} \sum_{j=1}^{N_y} (1-\alpha) |y_{true(i,j)}^t - y_{pred(i,j)}^t| & otherwise \end{cases} \quad (1)$$

Where $y_{true(i,j)}^t$ is the ice chart value for the pixel (i,j) at time t, and $y_{pred(i,j)}^t$ is the predicted value from the model for the pixel (i,j) at time t. $N_x\ and\ N_y$ are the number of cells in the x and y direction respectively, and $N_t$ is the number of time steps.

The final model architecture chosen from the hyperparameter optimization process is shown schematically in Figure 3. The model comprises one ConvLSTM layer, five hidden CNN layers, and one output CNN layer. The ConvLSTM layer employs 16 filters, while the CNN layers utilize 32, 64, 32, 16, 8, and 1 filters for the five hidden layers and the output layer, respectively. The ConvLSTM layer uses a 1x1 kernel size, while the CNN layers use a 3x3 kernel size. The optimum loss function parameter α was found to be 0.65, and the optimal lookback period was determined to be 15. A dropout ratio of 0.3 for the linear transformation of the inputs in the ConvLSTM layer was used. The ConvLSTM layer employed the Tanh activation function, whereas the CNN layers utilized the ReLU activation function. Additionally, the ConvLSTM layer succeeded with batch normalization and Tanh activation function while each of the inner CNN layers succeeded with batch normalization and Parametric ReLU (PReLU) activation function.

## Model performance assessment

To quantify model performance for optimization and comparison, we employed several metrics including the Root Mean Square Error (RMSE), simulation bias, and the correlation coefficient. All performance metrics presented herein were calculated based on the validation and testing datasets (2010-2018 and 2018-2022, respectively).

The assessment of the model performance included the calculation of two distinct RMSE values: one for the lake-wide average ice cover and another for pixel-by-pixel evaluations. The latter evaluation aimed to gauge the model's local accuracy in predicting ice cover, and spatial patterns in the model accuracy (hereinafter referred to as Local RMSE).

A comparison baseline model was developed by computing the average ice cover concentration for each day between November and April across all years in the dataset. This baseline model served as a reference point for comparison, to see if the model predictions represented improvement over a prediction based solely on the average ice cover chronology for the lake.

In addition to the model performance assessment mentioned above, six nearshore locations in Lake Michigan were selected to assess the model accuracy near important harbors and large cities: 1- Chicago, 2- Milwaukee, 3- Green Bay, 4- Port Inland, 5- Holland, and 6- Ludington. Ice predictions and ice charts values were extracted for these coastal locations using the closest landward cell in the grid to the location of interest (Figure 1).

## Results

The trained model was able to effectively simulate the Lake Michigan ice cover concentration. Figure 4 shows the monthly average simulated and observed ice cover for 2013-2014, which was a season of significant ice cover. The model can be seen to successfully replicate the spatial and temporal ice evolution starting from the shallow northern waters of Green Bay, as well as the progression of ice cover establishment from nearshore to offshore areas.

At the coarsest spatial scale, the lake-averaged daily ice cover chronology had seasonally-averaged RMSE ice cover values ranging between 0.014 for the low ice winter of 2019-2020 and 0.096 for the high ice winter of 2013-2014 (Figures 5-7, Table 2). The model bias over the training and validation periods is negligible on average, with a maximum magnitude of 0.016. In general, the model accuracy is reduced as the magnitude of the ice cover increases (Figures 6 and 7) but has a high overall correlation for all data points (r = 0.95) and negligible overall bias.

Figure 8 shows the starting and ending dates of the ice cover for the model and the ice charts for different years in the validation and testing dataset. The starting date is defined as the first day with an average ice cover concentration of more than 1% and the ending date is defined as the last day with an average ice cover concentration of 1% in the ice season. The model shows good skill in detecting the starting and ending dates of the ice cover for most of the years in the validation and testing datasets. The difference between the ice starting dates calculated from the model and the ice chart ranges between 0 and 11 days with a median of 4 days for the validation and testing datasets, while the difference between the ice ending dates ranges between 1 and 18 days with a median difference of 3 days. The model for most of the years in the validation and testing datasets starts the ice season later in the year than the ice charts data, but there is no consistent bias for the simulated ending date.

Annual averages of the ice cover spatial distribution show that the model can successfully replicate the average annual spatial pattern of ice cover in Lake Michigan, which generally has high ice cover towards the north of the lake, particularly in Green Bay and Traverse Bay (Figure 9). As described earlier for the daily (spatially-averaged) ice cover, the mismatch between the model and the observations is seen to be largest when and where the ice cover magnitude is also largest, with a slight consistent overprediction of ice in Green Bay as well as the southern shore of Lake Michigan, along the Indiana coastline.

The results of the model predictions for the six selected nearshore locations around Lake Michigan are shown in Figure 10, with site-specific statistics provided in Table 3. The model predictions matched the ice chart data with good accuracy for most of the locations except for Ludington where the model predictions underestimated the ice cover for most of the years on the testing dataset. The RMSE of the model's predictions at these locations varies between 0.14 and 0.21, the bias varies between -0.001 and 0.037, and the correlation coefficient ranges between 0.76 and 0.92. In general, the accuracies of the lake-averaged ice concentrations were significantly greater than the local predictions at individual locations (Figure 7; Tables 2 and 3).

Figure 11 shows the spatial distribution of the local monthly Normalized RMSE (NRMSE) for 4 years from the validation and testing datasets, which was calculated by dividing the RMSE by the range of errors for each pixel for the period of interest. The results show a positive correlation between the ice cover concentration and NRMSE, where areas with large ice cover have a relatively larger NRMSE. For example, shallow and near shore areas, that usually have relatively higher ice concentration, show relatively higher NRMSE. The results also show relatively large NRMSE in areas with large spatial gradients in the ice cover concertation, where the model tends to produce smoothed predictions (Figures 4 and 11).

Generally, the model captures the low-frequency variation of the lake-averaged seasonal ice cycle in terms of the ice concentration timing and magnitudes, while lacking some of the finer temporal structure on timescales of weeks or less (Figures 5 and 12; Table 2). The daily model-simulated ice cover peaks were usually delayed by less than one week relative to the observed peaks, especially for the highest ice seasons (seasons 2013-2014 and 2014-2015) (Figure 5).

Perhaps unsurprisingly, averaging the ice cover on weekly and monthly timescales improves both the lake-wide and local errors (Figures 12, 13). Monthly lake-wide ice cover values for the model had RMSE of 0.036 and 0.017 for the validation and testing datasets, respectively. Weekly whole-lake ice cover values had only slightly higher RMSE values of 0.043 and 0.023, respectively. Additionally, the maximum local RMSE decreased significantly from more than 0.4 for the daily predictions to less than 0.2 for the monthly predictions.

## Discussion

In this study, a new deep learning model was introduced for predicting the ice cover concentration in Lake Michigan, demonstrating significant proficiency in characterizing the lake-averaged lake ice concentration on daily, weekly, and monthly timescales. The RMSE for the average lake ice concentration values is comparable to or less than the 10% accuracy of the ice chart values themselves for all of the lake-averaged model results presented.

Discrepancies between the proposed model results and ice chart data can be attributed to several factors. Ice chart values are categorized with 10% intervals from 1983 onwards, and this inherent uncertainty in the ice concentrations, which are used as training data, translates to model uncertainties. Temporally, the daily ice chart data must often be interpolated between observation days, introducing potential data smoothing effects. Additionally, a bias exists for earlier ice chart data for the ice start date, since ice observations did not typically begin until after ice was established. Accordingly, the starting date of the ice cover in the historical data is usually earlier than the first observation recorded for a given season, which may have affected the model training. Resampling data before 2008, from 2.55 km to 1.8 km, may also impact accuracy, especially regarding small-scale spatial variations. Additionally, the ERA5-Land reanalysis product used as forcing for the model is not without errors and these errors cascade to model inaccuracies (Huang et al., 2021).

Although adding the shoreline proximity and water depth significantly improved the results, nearshore and shallow areas still experience higher NRMSE (Figure 11). One potential reason for that can be the slightly worse accuracy of the ERA5-Land in the coastal areas (Zou et al., 2022). Additionally, as mentioned earlier, the present model lacks a companion hydro- and thermo-dynamic model for water circulation and temperatures, instead relying meteorological variables as surrogates. The lack of a lake model leads reduces model complexity, but this reduction in complexity may result in a loss of direct accounting for potentially relevant processes such as circulation and mixing, wind waves, river inflows, and particular air-sea heat fluxes.

Despite these limitations, the presented DL model accuracy is comparable to the best currently available, physics-based model produced by NOAA's Great Lakes Environmental Research Lab (GLERL) presented in (Anderson et al., 2018). When comparing the two models for years 2015, 2016, and 2017, the DL model improves upon the GLERL model in the lake-wide RSME by scoring 0.039 in comparison with 0.05 for the GLERL model for the same period from January $1^{st}$ to May 31. However, the local RMSE for the DL was marginally worse scoring 0.127 compared to 0.12 for the GLERL model for the same period.

While the DL model can achieve similar or better accuracy for ice cover prediction compared to the physics-based model, both approaches have their own usage. The physics-based approach offers a strong foundation grounded in fundamental principles of ice formation and dynamics, providing interpretability and insight into the underlying physical processes, while the DL models have a black-box nature, lacking transparency and interpretability, which make the physics-based model better for understanding the underlying mechanisms driving ice concentration dynamics. Furthermore, the DL models typically rely on large volumes of labeled data for training, while the physics-based models usually need much less data for calibration, which gives the physics-based model an edge in remote areas. However, the physics-based models often require extensive computational resources, while the DL model, once trained, can be much faster, which gives the DL model an edge in long-term ice forecasting and climate change impact assessment. Moreover, the physics-based models are restricted to our understanding of the physics and the physical process included in them, while the DL models learn directly from the data, which makes the DL models better for simulating phenomena with poor physical understanding.

As seen in Figure 10, the model's performance varies at different selected locations. To enhance the model's accuracy at specific locations, one approach involves fine-tuning the model to prioritize accuracy in those areas. This adjustment can be achieved by modifying the loss function to one that is customized for selected locations. While this specialized model may not perform optimally across all locations, it should excel in the areas of interest.

This approach was applied to the chosen nearshore locations. However, the improvement in the area-specific models was not significantly better. For instance, the Chicago-tailored model resulted in improved RMSE values on both the validation and testing datasets from 0.173 to 0.165. However, for areas like Port Inland and Ludington, the tailored model did not show improvement over the original model. The failure of a customized, location-specific model optimization to produce better local predictions suggests that the bottleneck for higher-accuracy local predictions is either a flaw in the model architecture (e.g. unresolved processes) or in the data used to force and train the model.

The choice of the customized loss function was decided after experimenting with several alternative loss functions, and may prove useful in other DL models of lake processes and variables. Initially, the Mean Square Error (MSE) of the ice cover concentration was employed as the primary loss function. However, it was observed that the MSE tended to yield smoothed results, a characteristic inherent to this loss function (Thomas, 2020). Moreover, models trained using the MSE exhibited an unacceptably high bias, resulting in the underestimation of peak ice cover concentrations. This bias issue was particularly prominent due to the limited number of events with high ice cover values in the training dataset. Consequently, the customized loss function, as outlined in the methods section (equation 1), which penalizes underestimation more than overestimation was adopted. This decision led to a substantial enhancement in the model's performance, mitigating the bias and significantly improving the accuracy of ice cover predictions.

Substantial improvements were evident when analyzing weekly and monthly model output in comparison to daily model output. This improvement can primarily be attributed to the occasional disparities that arise between the starting and ending ice data (Figure 7), as well as disagreements concerning peak ice days (Figures 5 and 7). The act of averaging the results over weekly or monthly intervals effectively mitigated these discrepancies, which resulted in more accurate weekly, monthly, and seasonal data (Figures 12 and 13). Depending on the application for which the ice model is being used, accurate weekly- and monthly-averaged concentrations may be acceptable in lieu of detailed daily predictions.

Accepting the model as valid, hindcasts of Lake Michigan ice cover concentrations can be extended to cover the historical period spanned by the forcing (reanalysis) data. This extended hindcast, in turn, can then be used to better understand interannual and long-term ice cover trends, relationships with changing climatic forcing, and potential effects on the lake ecosystem, coastal processes, marine transportation, and more. These applications usually use monthly or seasonally averaged metrics of the ice cover concentration, which can be obtained from the current model with reasonable accuracy. The application of the model to produce such a hindcast is the subject of ongoing work by the authors.

## Conclusion

In this study, a new deep learning model, based on ConvLSTM and CNN, is introduced for hindcasting the ice cover concentration in a large lake, Lake Michigan (USA). The model utilized inputs from the ERA5-Land weather dataset, along with water depth and shoreline proximity data. Training, validation, and testing of the model were conducted using NOAA ice chart datasets spanning from 1973 to 2022. The application of ensemble learning techniques and the introduction of a customized loss function were applied to mitigate the model bias and enhance its generalization.

The model simulated ice concentrations with good accuracy across various seasons and years, exhibiting high accuracy in lake-averaged ice concentrations, and slightly diminished accuracy for local ice cover values.  Additionally, the model accuracy is comparable to existing physics-based models for the lake-wide predictions.

In addition to the daily ice cover simulations, the model results were averaged on weekly and monthly timescales, which significantly reduced the model error, potentially enhancing its utility in various environmental and ecological applications, such as assessments related to fish recruitment, maritime transportation, coastal resilience, and algae bloom activity.

Overall, the proposed model illustrates the potential of deep learning-based approaches as powerful tools for ice cover simulation, making it a valuable asset for extending the Great Lakes ice cover concentration datasets, and potentially offers a new avenue for developing ice cover predictions and climatic assessments.

## Acknowledgments

This work was supported in part by the Illinois-Indiana Sea Grant College Program, grant number NA18OAR4170082, and the Indiana Department of Natural Resources Lake Michigan Coastal Program, grant number NA20NOS4190036. Hazem Abdelhady also acknowledges support from the Lyles School of Civil Engineering and The American Society of Civil Engineers (ASCE).

# Figures

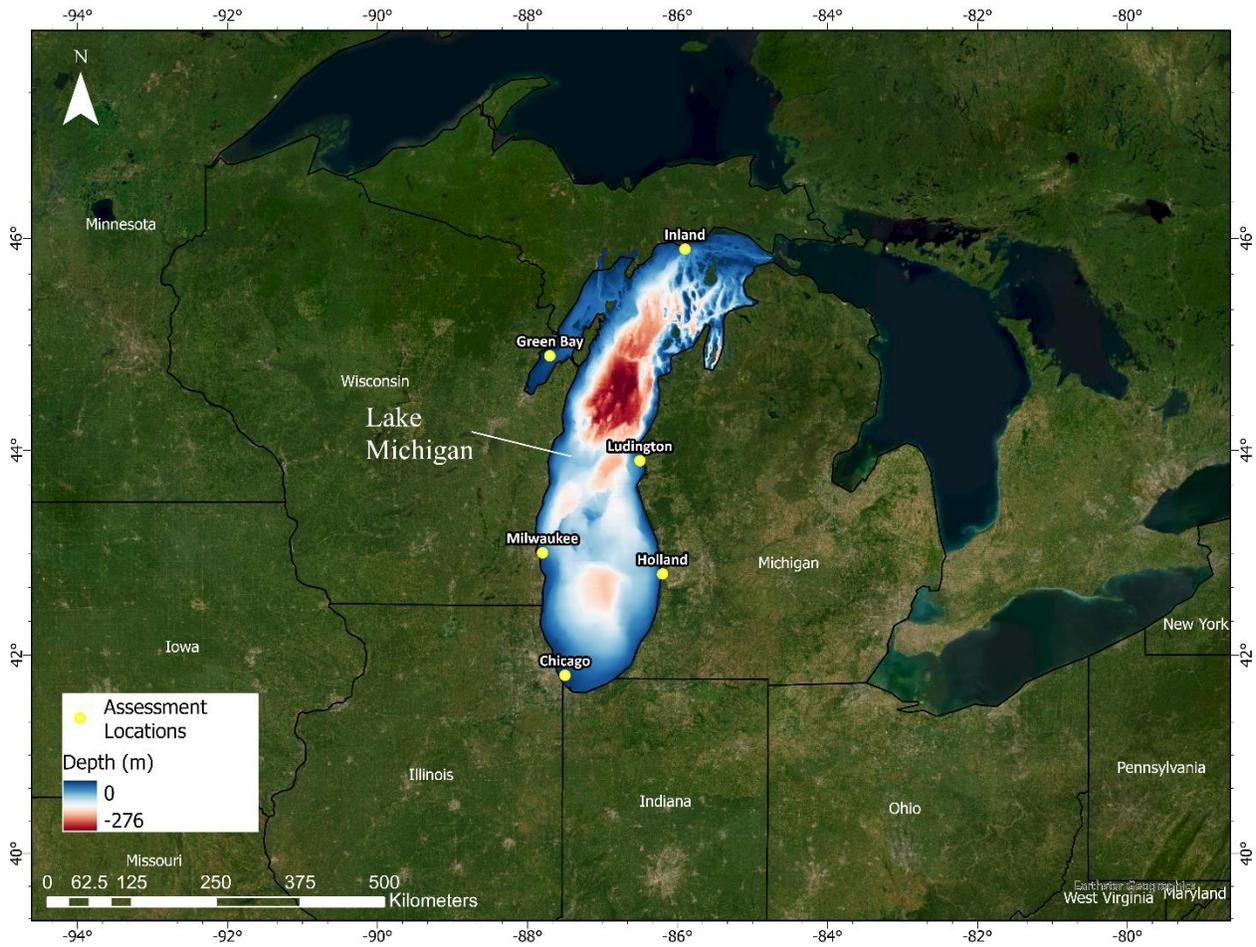

*Figure 1 Lake Michigan location and bathymetry. Ice concentration was simulated at all locations in the lake, but locations shown here were used for detailed model assessment. (Lake Michigan Mean Low Water datum is used for the depth)*

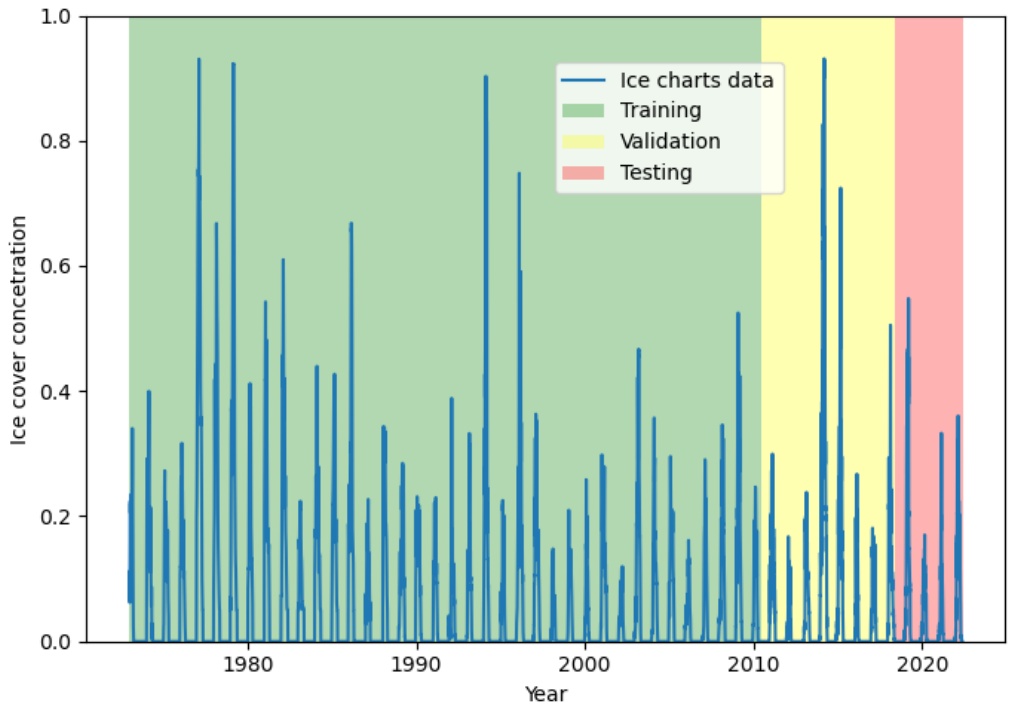

*Figure 2 Spatially-averaged Lake Michigan ice concentrations from 1973-2022* (NOAA, 2022). *Spatially averaged ice chart data showing the parts used for training, validation, and testing for the DL model.*

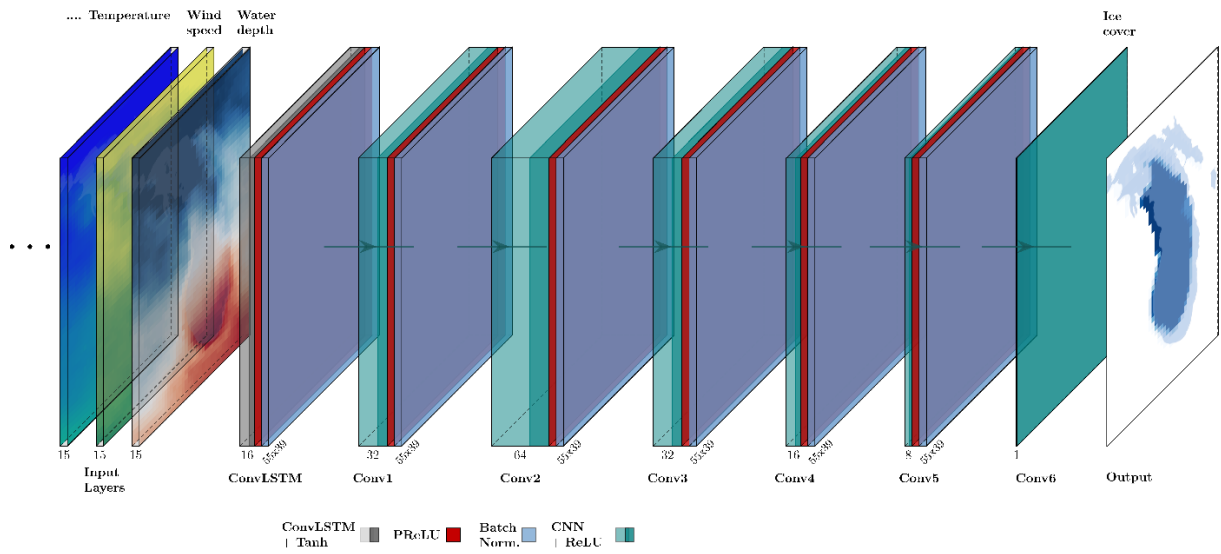

*Figure 3 The Architecture of the proposed deep learning model. The figure was created using the PlotNeuralNet template (https://github.com/HarisIqbal88/PlotNeuralNet/).*

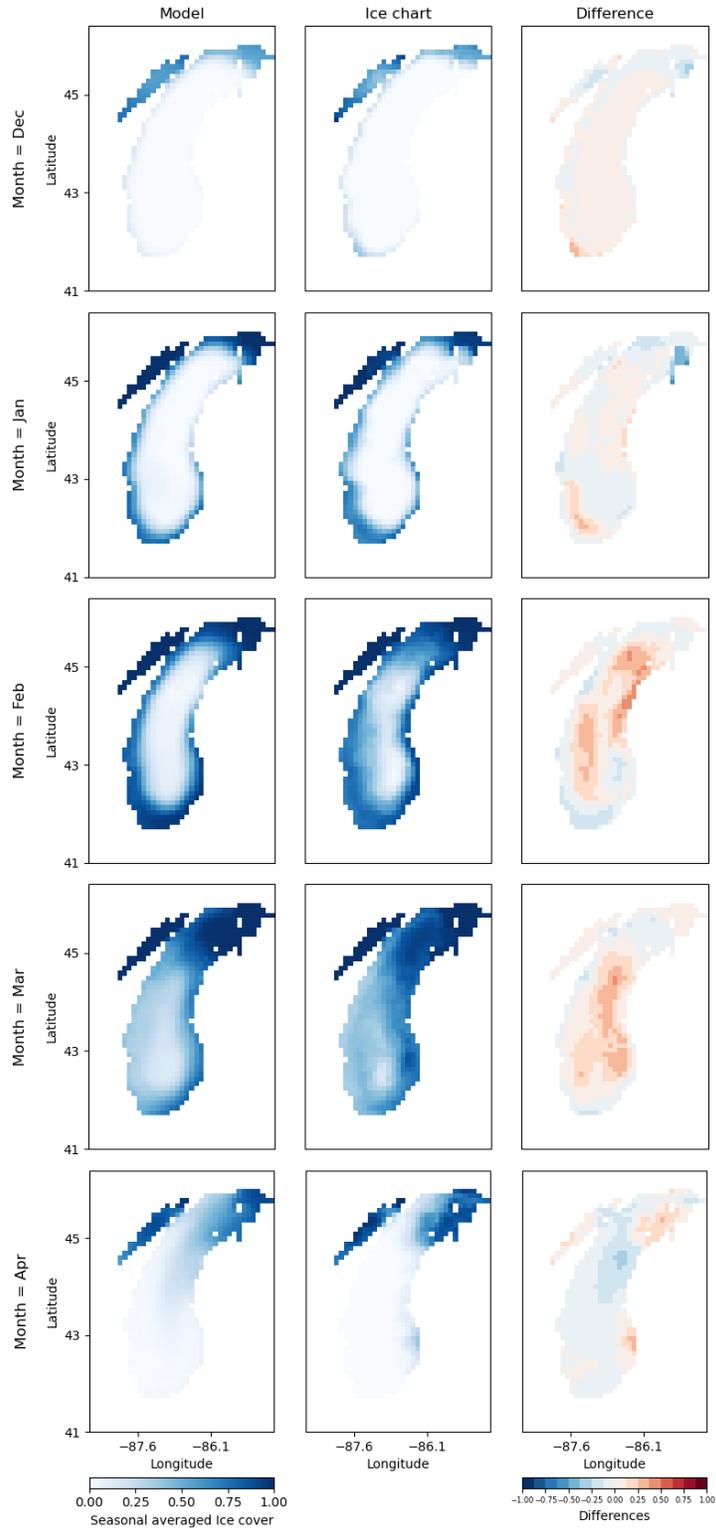

*Figure 4 Model predictions versus the ice charts for the seasonal ice cover evolution in season 2013-2014. Each row in the first two columns represents a monthly average ice cover for the model prediction and ice charts respectively. The third column represents the differences between the ice chart and model prediction.*

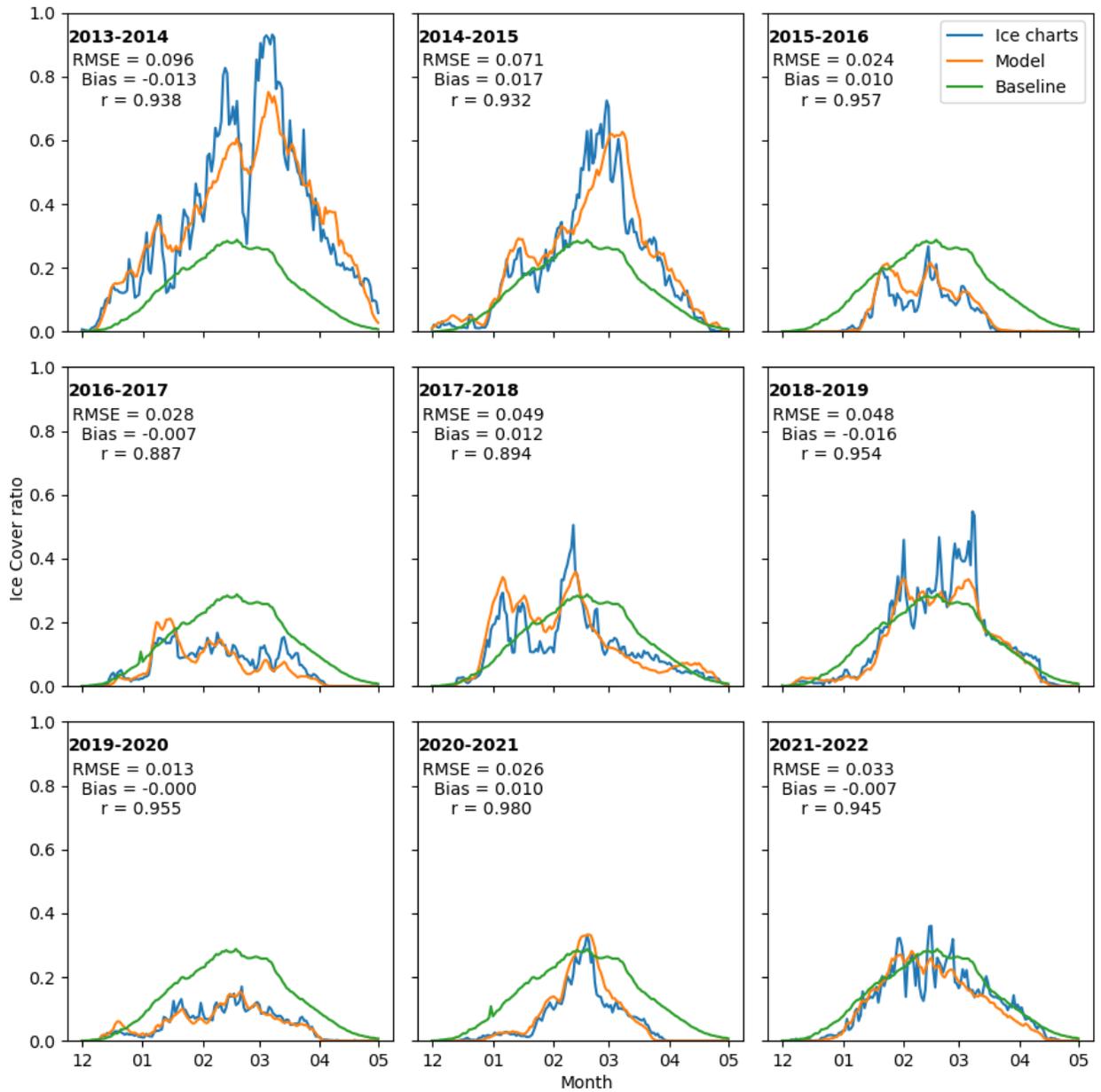

*Figure 5 Spatially-averaged daily ice cover simulated by the model for seasons from the validation and testing datasets. The baseline model shown is the average ice cover chronology calculated with the ice atlas for the years 1973-2022.*

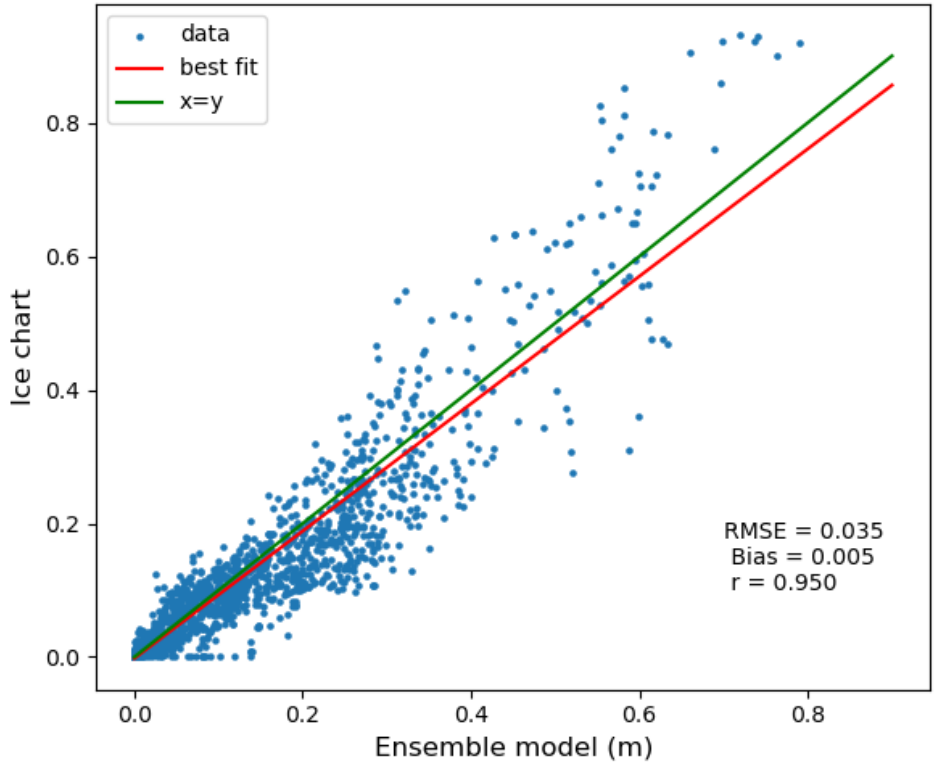

*Figure 6. Lake-wide daily model ice cover predictions relative to ice chart observations for all days during the validation and testing datasets.*

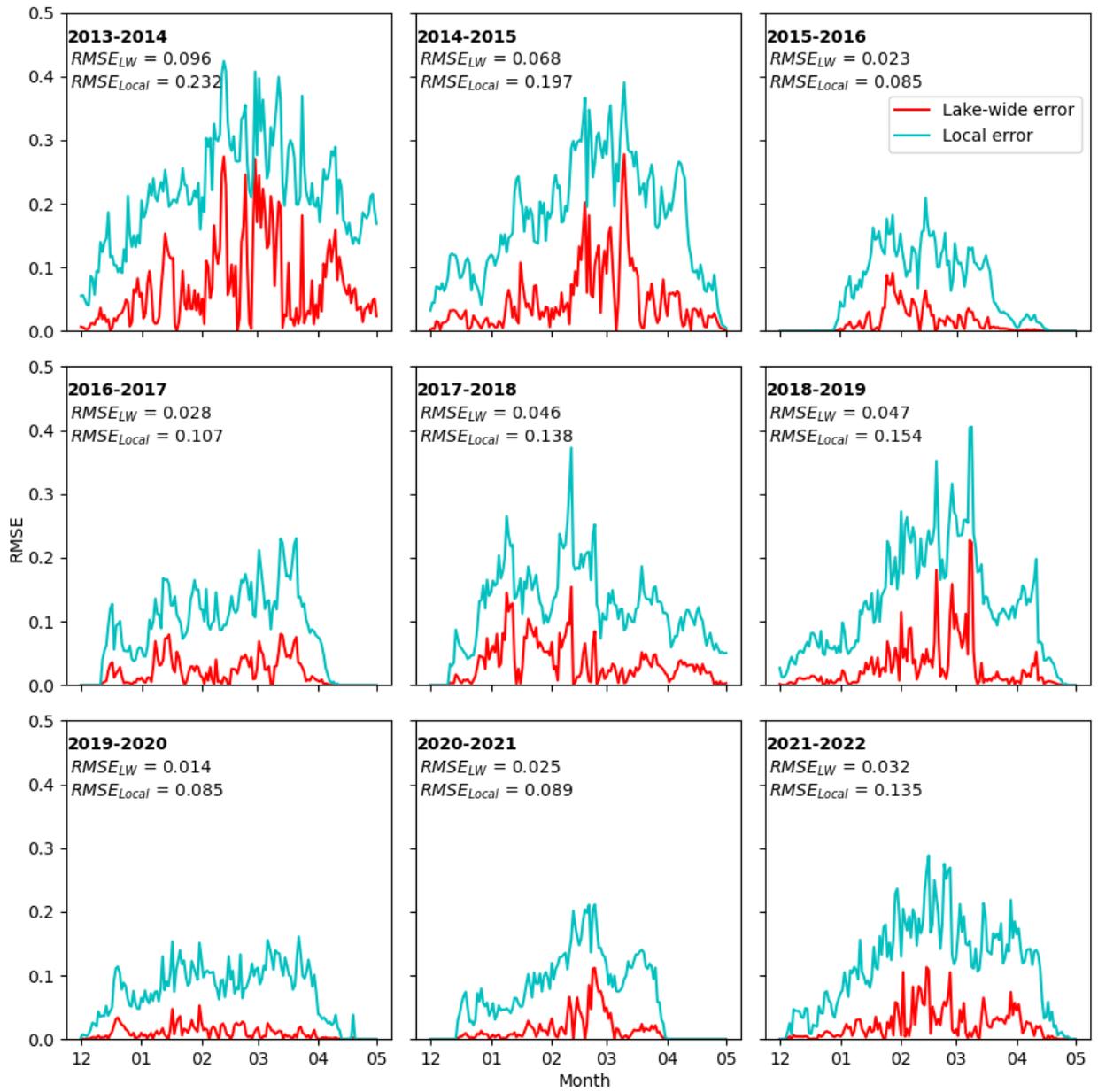

*Figure 7 Error in lake-averaged daily ice cover and average error in local ice cover predictions.*

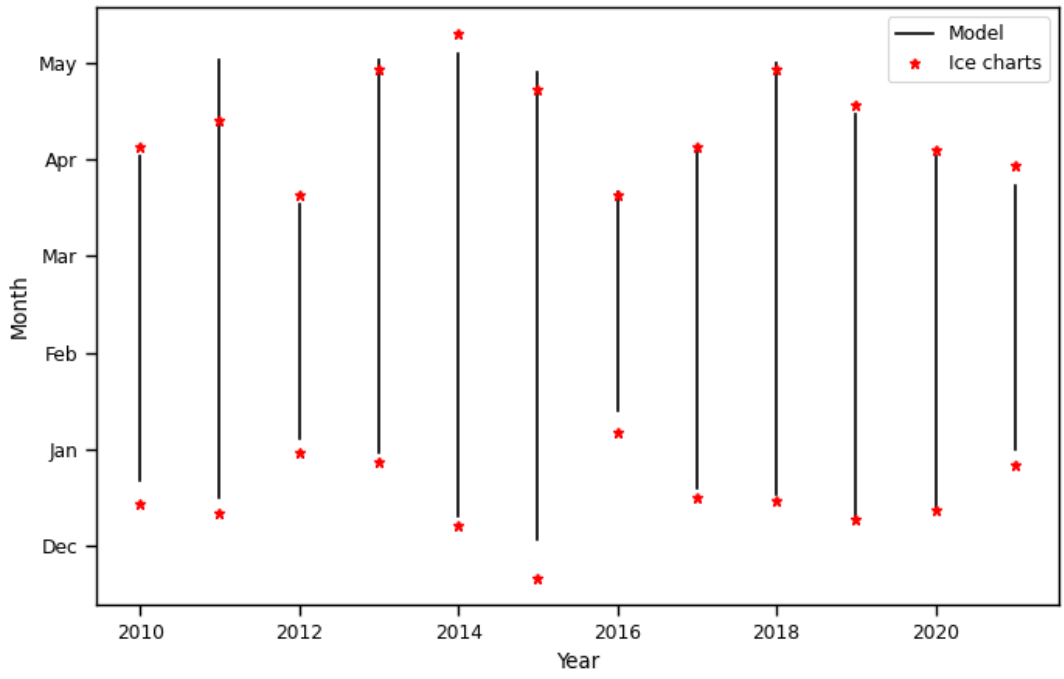

*Figure 8 Simulated and observed starting and ending dates for the ice seasons in the validation and testing datasets.*

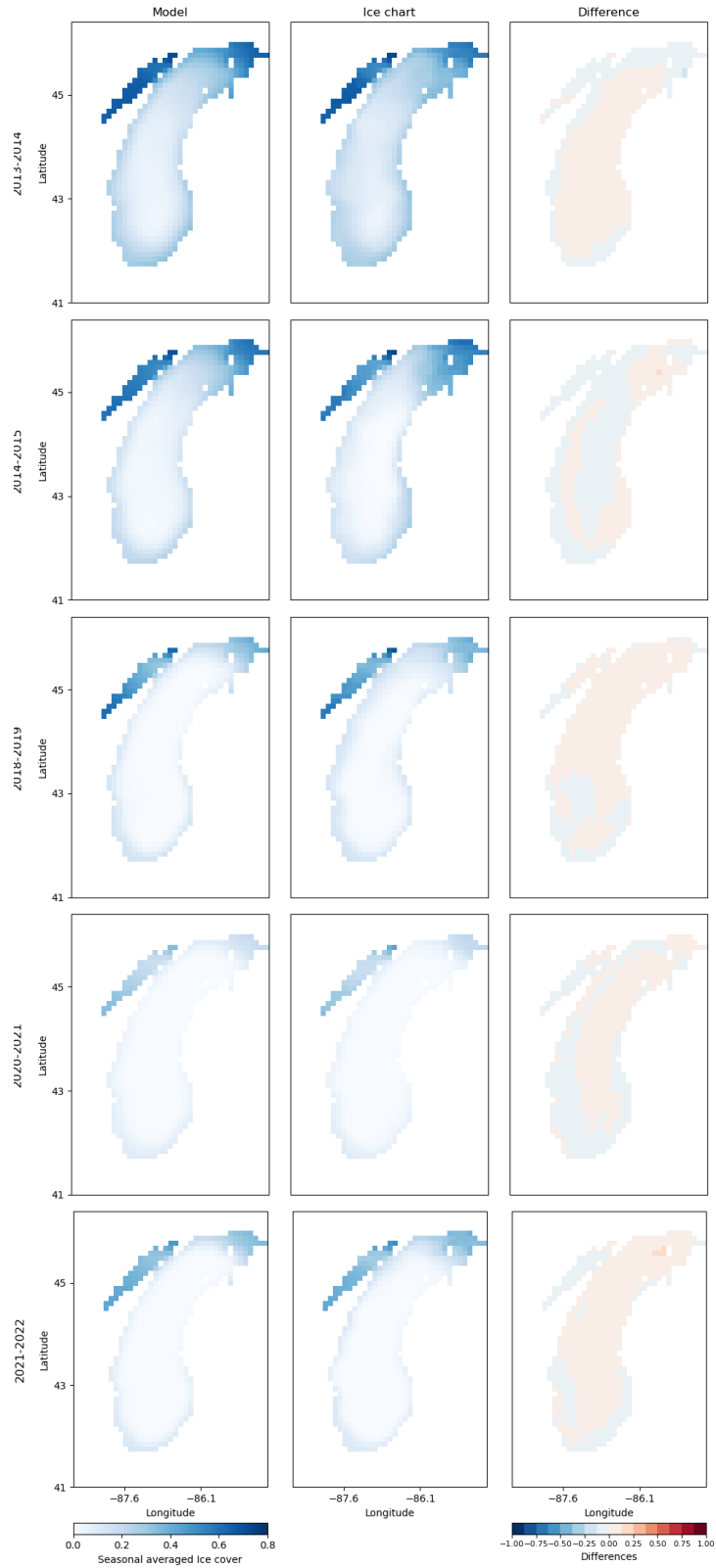

*Figure 9 Seasonally-averaged ice cover simulations (left), observations (middle), and error (right) for select simulation years during the training and validation years.*

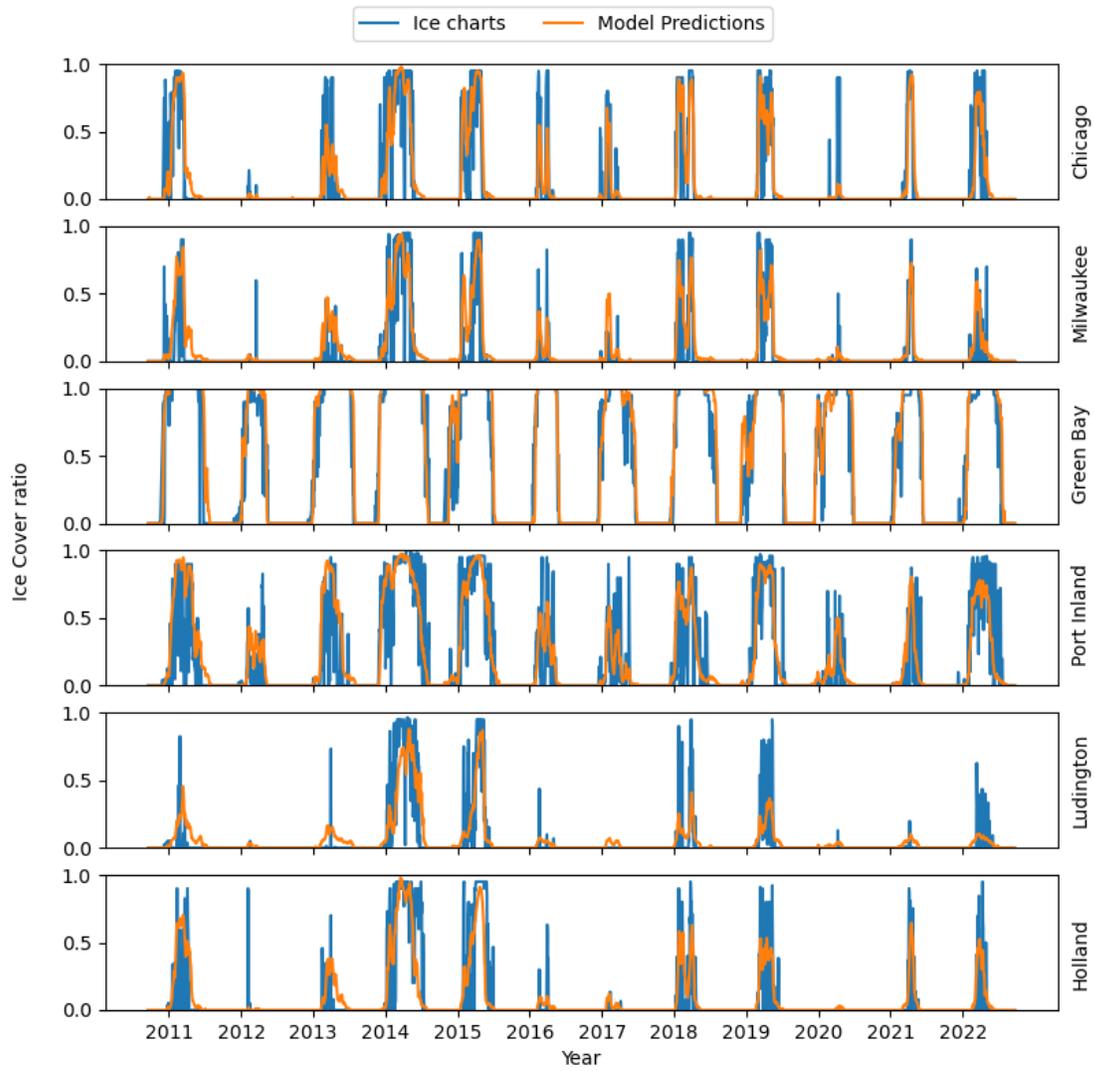

*Figure 10 Model prediction at the six selected nearshore locations in Lake Michigan, for the period November 1 to May 31 for all years in the validation and testing datasets.*

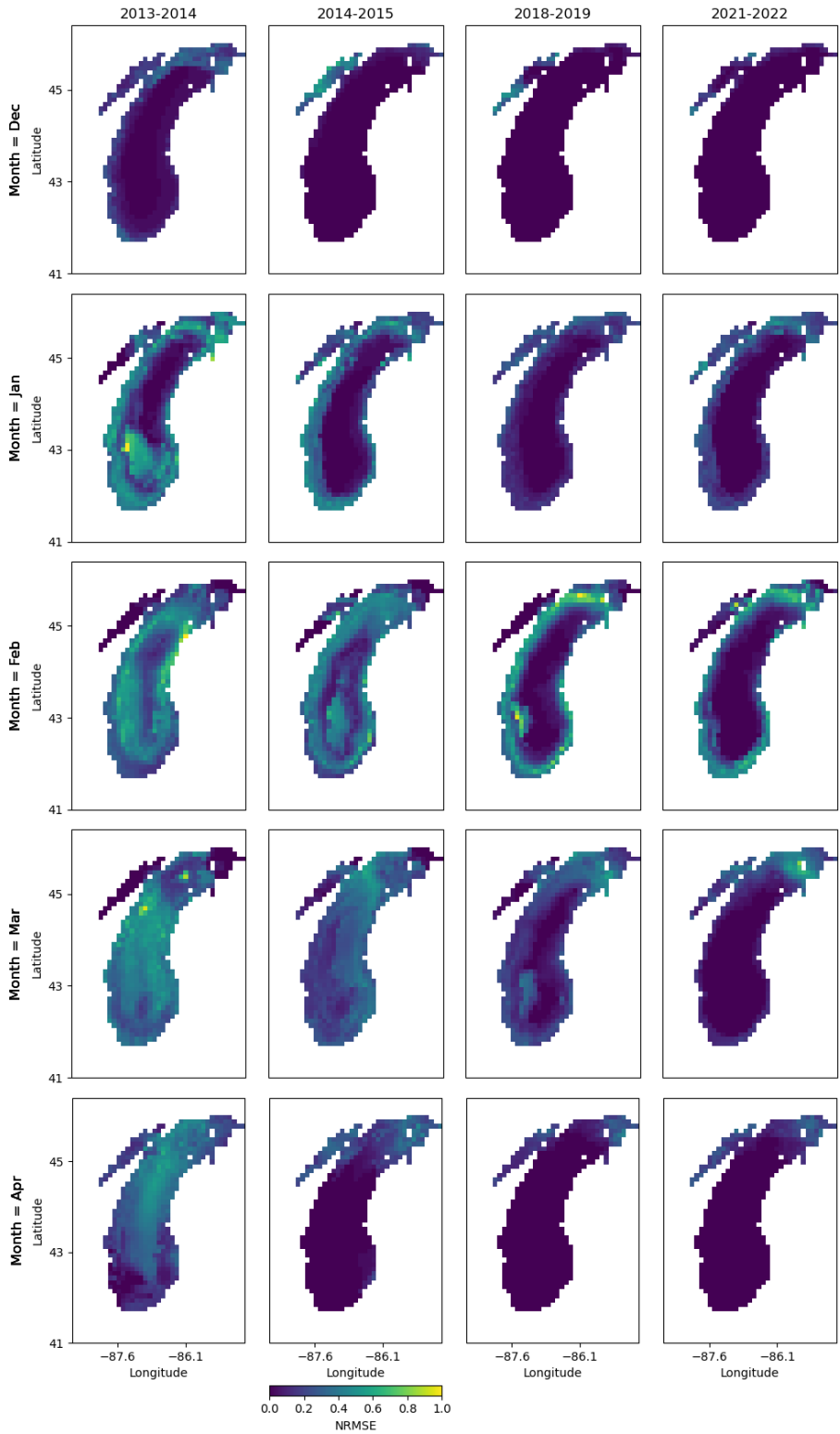

*Figure 11 Monthly averaged local NRMSE for 4 ice seasons from the validation and the testing datasets.*

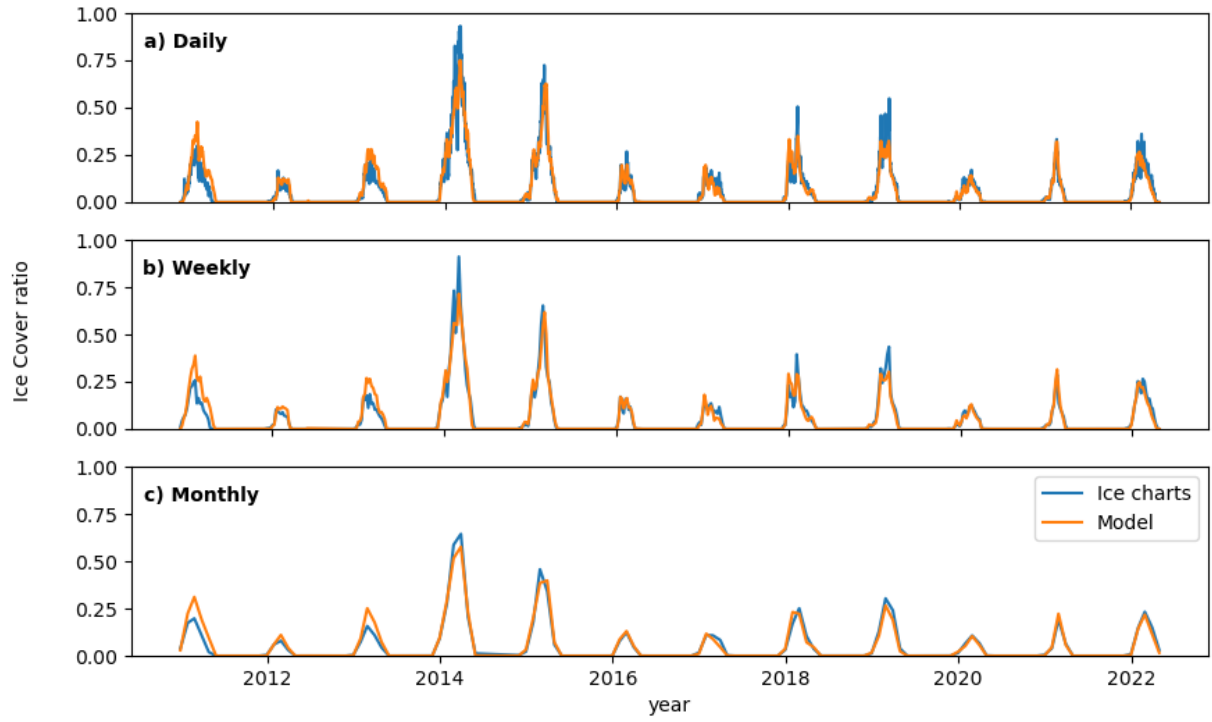

*Figure 12 Lake-averaged ice predictions and observations as (a) daily, (b) weekly-averaged, and monthly-averaged values*

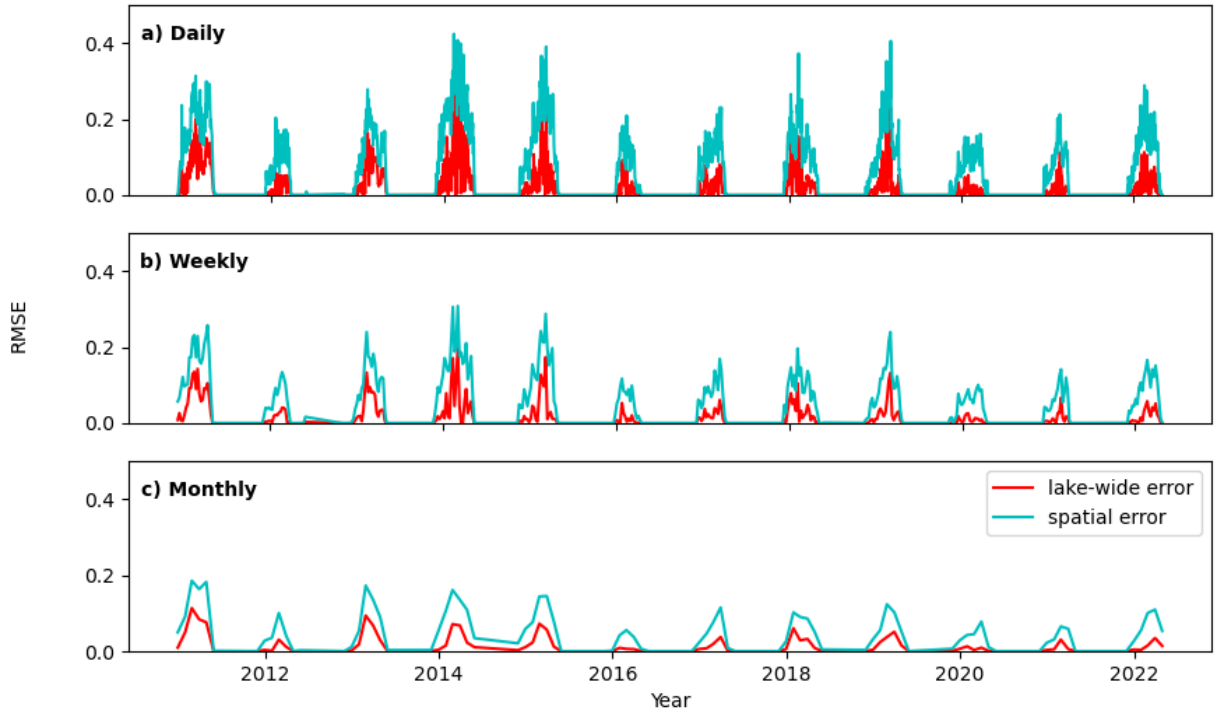

*Figure 13 a) Comparison between the errors for the predicted daily, weekly-averaged and monthly-averaged ice cover.*

*Table 1 The input features used in the model development.*

| Input features | Source |
|---|---|
| 2m air temperature | ERA5-Land |
| 30-day moving averages temperature | Calculated using the 2m air temperature from the ERA5-Land dataset |
| 150-day moving averages temperature | Calculated using the 2m air temperature from the ERA5-Land dataset |
| Net short-wave radiation | ERA5-Land |
| Wind speed | ERA5-Land |
| Wind direction | ERA5-Land |
| Lake water depth | Lake Michigan Bathymetry and Calumet Harbor, IL water level gauge. |
| Shoreline proximity | Calculated using USGS Great Lakes and Watershed Shapefiles (USGS, 2010). |

*Table 2 Average model assessment metrics for the validation and the testing datasets.*

|  | Validation | Testing |
|---|---|---|
| RMSE (daily) – Lake-wide | 0.049 | 0.029 |
| RMSE (daily) – Local | 0.132 | 0.102 |
| Bias (daily) | 0.008 | -0.008 |
| $r_{Pearson}$ (daily) | 0.95 | 0.96 |
| RMSE (weekly) – Lake-wide | 0.043 | 0.023 |
| RMSE (weekly) – Local | 0.103 | 0.073 |
| Bias (weekly) | 0.008 | -0.0067 |
| $r_{Pearson}$ (weekly) | 0.96 | 0.98 |
| RMSE (monthly) – Lake-wide | 0.036 | 0.0167 |
| RMSE (monthly) – Local | 0.081 | 0.054 |
| Bias (monthly) | 0.0082 | -0.007 |
| $r_{Pearson}$ (monthly) | 0.97 | 0.99 |

*Table 3 Performance metrics for the selected nearshore locations*

| Location | Latitude | Longitude | RMSE | Bias | $r_{Pearson}$ |
|---|---|---|---|---|---|
| Chicago | 41.9° | -87.5° | 0.173 | -0.001 | 0.83 |
| Milwaukee | 43.0° | -87.8° | 0.142 | 0.037 | 0.82 |
| Green Bay | 44.9° | -87.7° | 0.170 | 0.036 | 0.92 |
| Port Inland | 45.9° | -85.9° | 0.210 | 0.027 | 0.81 |
| Ludington | 43.9° | -86.5° | 0.129 | 0.003 | 0.80 |
| Holland | 42.8° | -86.3° | 0.172 | 0.007 | 0.76 |